\documentclass[fleqn,twoside]{article}

\usepackage{espcrc2}

\usepackage{epsfig}

\usepackage{amssymb}

\usepackage[figuresright]{rotating}

\newcommand{\AmS}{{\protect\the\textfont2

  A\kern-.1667em\lower.5ex\hbox{M}\kern-.125emS}}

\title{Proposal for a Hadron Blind Detector for PHENIX}

\author{
  A. Kozlov\address[WIS]{Weizmann Intitute of Science, 76100 Rehovot, Israel},
  C. Aidala\address[BNL]{Brookhaven National Laboratory, Upton, NY11973, USA}, 
  B. Azmoun\address[SB]{State University of New York, Stony Brook, NY11974, USA},
  Z. Fraenkel\addressmark[WIS],
  T. Hemmick\addressmark[SB],
  B. Khachaturov\addressmark[WIS],
  A. Milov\addressmark[SB],\\
  I. Ravinovich\addressmark[WIS],
  I. Tserruya\addressmark[WIS],
  S. Stoll\addressmark[BNL],
  C. Woody\addressmark[BNL] and
  S. Zhou\addressmark[WIS]
}

\begin{document}

\DeclareGraphicsExtensions{.jpg,.pdf,.mps,.png}

\begin{abstract}

  A Hadron Blind Detector (HBD) is proposed as upgrade of the PHENIX detector at
  RHIC, BNL. The HBD will allow the measurement of low-mass e$^+$e$^-$ pairs
  from the decay of the light vector mesons $\rho$, $\omega$, $\phi$ and the
  low-mass continuum in Au-Au collisions at energies up to $\sqrt{s_{NN}}$= 200
  $GeV$. From MC simulations and 
  general considerations, the HBD has to identify electrons with very high
  efficiency ($>$ 90$\%$), double hit recognition better than 90$\%$, moderate
  pion rejection factor of $\sim{200}$ and radiation budget of the order of
  1$\%$ of a radiation length. The first choice under study is a windowless
  Cherenkov detector, operated with pure $CF_4$,
  in a special proximity focus configuration with a $CsI$ photocathode and a
  multistage GEM amplification element. 

  \vspace{-1mm}

\end{abstract}                                

\maketitle
\section{Introduction}

The PHENIX experiment at RHIC, the Relativistic Heavy Ion Collider at Brookhaven
National Laboratory, is specially designed to measure  electrons, muons,
photons, neutral mesons and identified charged hadrons in ultra-relativistic
heavy-ion collisions. With this broad spectrum of observables PHENIX is able to
study in detail the collision dynamics and in particular the earliest
stages where the new state of matter, known as the Quark-Gluon plasma (QGP) is
expected to be formed.   

\begin{figure}[htp]
  \vspace{-5mm}
  \centering
  \epsfig{file=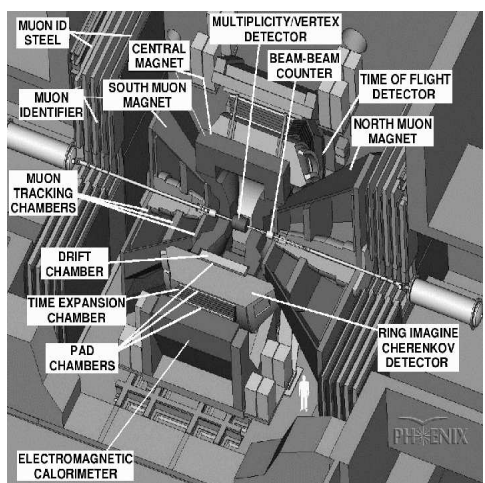,width=15pc, height=15pc}
  \vspace{-10mm}
  \caption{View of the PHENIX detector.}
  \label{fig:phnx_1}
  \vspace{-5mm}
\end{figure}

The layout of the PHENIX detector~\cite{phnx} is shown in
Fig.~\ref{fig:phnx_1}. It consists of two central spectrometer arms, each one
covering 90$^{\circ}$ in azimuth and $\vert\eta\vert<$ 0.35 and two forward
muons spectrometers with full azimuthal coverage in the pseudo-rapidity range
1.2 $<\vert\eta\vert<$ 2.2. The central arm detectors are located outside an
axially symmetric magnetic field provided by a central magnet. The detectors
include a high resolution tracking system consisting of drift chambers, pad
chambers, and time-expansion chambers. Together with a high resolution
time-of-flight hodoscope, electromagnetic calorimeters, and a gas-filled ring
imaging Cherenkov counter, the central arms provide excellent hadron, electron,
and photon identification capabilities over a wide range of transverse
momenta. The presently achieved particle momentum resolution is
$\delta p/p = 0.8\%\oplus1.0\%p$ $GeV/c$, close to the design value of $\delta
p/p = 0.6\%\oplus0.3\%p$ $GeV/c$.

For global event characterization, the setup
includes a silicon vertex detector, two zero-degree calorimeters sensitive to
neutrons emitted along the beam axis, and a set of beam-beam counters to
determine the reference for time-of-flight measurement and vertex
determination. The data acquisition system is capable of archiving 60 MB/s
which corresponds to about 400 minimum bias Au-Au events per second. 

The present set-up of the PHENIX detector does not allow a good measurement of
low-mass electron pairs with $m_{e^+e^-} \leq 1 GeV/c^2$. Tracks with momentum
$p\leq 200 MeV$ cannot leave the magnetic field region. Thus, very often for
the pairs originating from $\gamma$ conversions and $\pi^{0}$ Dalitz decays only one
of the partners is detected. This leads to a huge combinatorial background that
increases quadratically with the number of charged tracks N$_{ch}$.  At the high
multiplicities of RHIC, this makes signal detection extremely difficult. For
example, first PHENIX results show, in agreement with Monte Carlo studies, that
even with a large single electron $p_t$ cut of 300 $MeV/c$, the $\phi$ meson
measurement has a signal to background ratio, S/B, of about 1/20~\cite{dsm} and
at lower invariant masses $m_{e^{+}e^{-}}\approx400 MeV/c^{2}$, S/B$\sim$1/100. 


A proposal for an upgrade of the PHENIX detector to dramatically improve the
performance in the measurement of low-mass $e^+e^-$ pairs is presently under
study. The upgrade has two main elements: the installation of a second coil,
foreseen in the original PHENIX design, to generate an almost field-free region
extending up to $\sim{60}$ $cm$ in the radial direction and an additional detector in that region
to recognize, and provide the necessary rejection of, Dalitz and conversion
tracks. 

In this paper, we limit the discussion to a possible realization of this
additional detector. We first present the detector concept, expected performance
and benefits. We then describe the current R\&D program to address a number of
issues and questions raised.  

\section{General Concept of the Upgrade}

The basic idea of the proposed  upgrade exploits the fact that the opening angle
of electron pairs from $\gamma$ conversions and $\pi^0$ Dalitz decays is
small compared to the pairs from light vector mesons. In a field-free region
this angle is preserved and by applying an opening angle cut one can reject more
than 90$\%$ of the conversions and $\pi^{0}$ Dalitz decays, while coserving
most of the signal. 

Monte Carlo studies at the principle level~\cite{prop}, with an additional
detector in the zero field region, assuming perfect spatial resolution and
perfect electron identification give a S/B ratio of $\sim{10}$ at the $\phi$
meson mass region with an opening angle cut of $\sim{180}$ mrad. The simulations
show that the background tracks remaining after such cut are close to the
boundaries of the fiducial acceptance. Therefore adding a \textit{veto area} to
the inner detector further improves the S/B ratio. For example, increasing the
acceptance in the azimuthal direction from 90$^{\circ}$ to 120$^{\circ}$ and
from $\vert\eta\vert\leq$ 0.35 to $\vert\eta\vert\leq$ 0.50 in pseudo-rapidity allows us to reach a
S/B ratio of about 25~\cite{prop}. 

At this level of rejection, the quality of the low-mass $e^+e^-$ pair
measurement will not any longer be limited by the combinatorial background of
conversions and $\pi^0$ Dalitz decays but rather by the background originating
from semi-leptonic decays of charmed mesons. Reducing this latter source is
beyond the scope of the proposed upgrade. Current crude estimates of the S/B
including the charm background give a S/B $\sim{1}$.

\section{Hadron Blind Detector}

The results of the simulations and general considerations allow us to determine the
inner detector specifications: electron identification  with an efficiency $> 90
\%$ including double hit recognition; a moderate $\pi$ rejection factor as low
as 100-200; the detector must fit within the radial distance 20$\leq$ r $\leq$70
$cm$; the detector should cover a slightly larger acceptance (\textit{veto
  area}) than the central arms; a radiation budget of the order of 1\% of a radiation length.

The requirements on electron identification limit the choice to a Cherenkov type
detector. After careful consideration of relevant options (configuration,
radiator gas, window, detector gas, photocathode, amplifying element, readout
scheme) our first choice is an HBD (Hadron Blind Detector)~\cite{hbd1,hbd2} in the following
configuration: a Cherenkov detector operated with pure $CF_{4}$ (or a
$CF_4$ based mixture) both as radiator and detector gas, in a special windowless
proximity
focus geometry, with a $CsI$ photocathode, multistage GEM (Gas Electron
Multiplier)~\cite{gem0,amos} as detector element and pad readout.

In this configuration, Cherenkov photons from an electron passing through the
radiator are directly collected on the $CsI$ photocathode, evaporated onto the
first GEM, forming a circular blob, not a ring as in a RICH detector.

\begin{figure}[htp]

  \vspace{-5mm} 

  \centering
  
  \epsfig{file=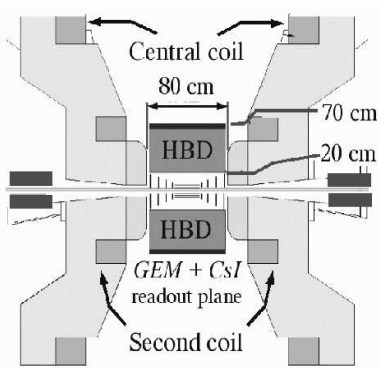,width=12pc, height=12pc}

  \vspace{-10mm}
  \caption{The HBD location in the PHENIX detector.}
  \label{fig:hbd1}
 \vspace{-10mm}
\end{figure}

\begin{figure}[htp]
  \vspace{-5mm}
  \centering

\epsfig{file=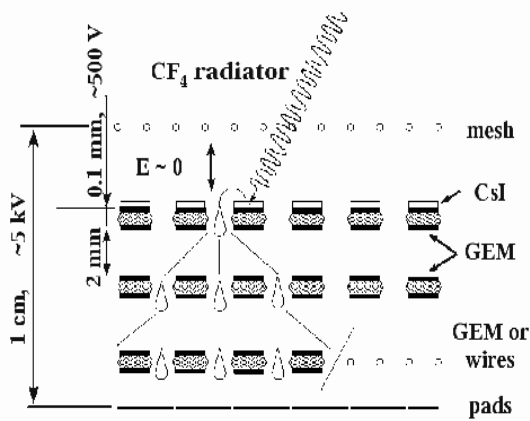,width=17pc, height=14pc}

  \vspace{-12mm}
  \caption{Layout of the readout element of the HBD.}
  \label{fig:hbd2}
  \vspace{-8mm}
\end{figure}

The location of the HBD in the PHENIX detector is shown in Fig.~\ref{fig:hbd1}
and a possible layout of the readout element is shown in  Fig.~\ref{fig:hbd2}.

This scheme exhibits a number of very attractive features.
The use of a windowless detector with $CF_4$ results in a very broad bandwidth
between the $CsI$ threshold of $\sim{6} eV$ up to the $CF_4$ cut-off of
$\sim{11.5} eV$
and a large figure of merit $N_0 \sim{940} cm^{-1}$ ($N_0 = \int Q.E.\,dE$). After including losses
incurred by the optical transparency of the entrance mesh (10\%) and the
photocathode (25\%) a very large number of detected photoelectrons, $N_{pe} =
40$, is expected in a 50 $cm$ long radiator. This large number of photoelectrons
ensures a very high electron efficiency. The ''reflective'' photocathode
scheme makes the  $CsI$ totally screened from photons produced in the avalanche.
The mesh before the first GEM is needed to ensure a zero field above the
photocathode, which is best for electron extraction in the ''reflective'' mode
\cite{amos}. Since the pattern produced by the detector is a circular blob
(maximum radius of 1.8 $cm$ in a 50 $cm$ long radiator) there is no a priori need to detect
single photo-electrons as in a RICH detector. This has the double benefit of 
low-granularity (the pad size can be comparable to the blob size) and a relatively low 
gain operation (since a pad will collect a relatively large primary charge). 

\vspace{-3mm}

\section{Expected Performance of the HBD}

\subsection{Occupancy and event rate}

 The occupancy and interaction rate expected at the HBD are very
modest. With  a charged particle density dN$_{ch}/d\eta = 650$ in central Au-Au
collisions at $\sqrt{s_{NN}} = 200 GeV$ the largest detector occupancy is
calculated to be 0.02 charged particles/$cm^{2}$.
For minimum bias events the occupancy drops by approximately a factor of $\sim{3}$
The expected interaction rate is 1400 Hz at the design RHIC luminosity. 

\subsection{Scintillation of $CF_4$}

$CF_4$ is known to scintillate and its scintillation properties were studied by
several groups~\cite{cf41,cf42}. The main feature of the scintillation spectrum
is a line at 163 $nm$ where the $CsI$ is sensitive.

Since scintillation occurs uniformly in $4\pi$ whereas the Cherenkov light is
emitted in a very narrow cone a considerable reduction of the scintillation
background can be achieved by installing shades. With a simple simulation, the
scintillation background over the size of a blob is found to be 4
photo-electrons. Adding radial shades close to the photocathode, 5 $cm$ high and
with 5 $cm$ spacing in both directions, reduces the background by a factor  of
three while reducing the Cherenkov photons by less than 10\%.  

\subsection{Detector response to electrons and hadrons}

The detector performance for a few options with $CF_4$-based gases is shown 
in Table ~\ref{table:1}. (More options are considered in \cite{prop}).

The table shows the estimated response to electrons and hadrons. The response to
electrons is characterized by the number of detected photoelectrons $N_{pe}$,
the maximum radius of the blob $R_{blob}$ and the double hit recognition
$DHR$. $DHR$ is the probability to resolve two electrons with zero opening angle
using amplitude analysis, and accepting 5$\%$ misidentification of single electrons
as double hits. 

The response to hadrons is characterized by the number of
electrons $N_e$ produced by a hadron within the size of a blob. It may be
uniformly spread over the whole detector area as in the case of scintillation
(G) or localized (L) in one pad when it results from ionization. For the latter
we have assumed that  the relevant ionization path-length for a particle
traversing the first GEM is only the distance above its surface corresponding to
a half of the hole pitch ($\sim{100}\mu m$) where the electric field collects the
primary ionization inside the holes. 

The last column of  Table ~\ref{table:1} shows an estimate of the $\pi$
rejection factor that can be achieved. It is defined as the number of incident
pions divided by the number of $\pi$ producing a signal above the lowest 5$\%$
of a single electron signal. 

\begin{table*}[htb]
  \centering
  \caption{The detector performance for different configurations and gas choices.}
  \label{table:1}
  \newcommand{\m}{\hphantom{$-$}}
  \newcommand{\cc}[1]{\multicolumn{1}{c}{#1}}
  \renewcommand{\arraystretch}{1.2} 
  \begin{tabular}[t]{|c|c|c||c|c|c||c||c|}
    \hline 
    \hline 
    \multicolumn{3}{|c||}{} & \multicolumn{4}{|c||} {Response to:} &{}\\
    \cline{4-7}
    \multicolumn{3}{|c||}{} & \multicolumn{3}{|c||} {electrons} & hadrons &{}\\
    \hline 
    Radiator gas & $\gamma_{th}$ & Shades & $N_{pe}$ per e$^{\pm}$ &
    R$_{blob}$ [$cm$] & DHR[$\%$] &$N_{e}$ G/L 
    & {$\pi$ rejection} \\
    \hline
    $CF_4$ & 28 & No & 40 & 1.8 & $>$90 & 4/1 
    & $>$10$^4$ \\
    $CF_4$ & 28 & Yes & 35 & 1.8 & $>$90 & 1/1 
    & $>$10$^4$\\
    $CF_4$/$Ne$ 50/50 & $\approx{40}$ & No & 30 & 1.3 & $\approx{90}$ 
    & 2/1 
    & $\approx{10^4}$ \\
    $CF_4$/$Ne$  10/90& $\approx{70}$ & No & 20 & 1.0 & $\approx{70}$ 
    & 1/1 
    & 350\\
    \hline
    \hline
  \end{tabular}

\end{table*}

\section{R$\&$D Program and plans}

Among the options listed in Table ~\ref{table:1}, the simplest and most attractive
is the first one. It gives around 40 photoelectrons per electron. It has excellent
$DHR$ and very high $\pi$ rejection. Recently the operation of a detector consisting 
of $CsI$ photocathode, three stages of GEM in pure $CF_4$ has been
demonstrated~\cite{gem4}. It gives some support that the proposed configuration
of the HBD can be realized. We consider this as our prime choice and the other 
options in the table and others listed in \cite{prop} are alternatives
to cure possible problems which may arise. Indeed there are a number of issues
and questions which need to be investigated in a comprehensive R$\&$D program which
has recently started.

The main goal of the R$\&$D program is to demonstrate the feasibility of the
proposed Hadron Blind Detector configuration. The specific issues to be
investigated are: 

$\Diamond{}$ Further studies of the $CF_4$ scintillation. We plan to measure the 
  scintillation signal produced by a MIP in the laboratory using cosmic muons. From our current 
  estimates the scintillation background is negligible. However, if needed it could 
  be substantially reduced by installing shades. Detailed simulations will be needed
  to optimize the shade configuration.

$\Diamond{}$  Study of aging effects on the GEMs elements and the $CsI$ photocathode in
  pure $CF_4$ atmosphere. This is one of the most critical issues. The $CsI$ QE degrades
  with a high dose of accumulated charge~\cite{qe}. From published
  numbers~\cite{qe}, this does not seem to represent a problem. The total charge
  expected to be accumulated on the $CsI$ over 10 years of continuous operation of the
  detector at design luminosity is well below 1mC/$cm^2$ (even assuming 100$\%$
  ion feedback). However, aging in the presence of $CF_4$ might be more severe
  and needs to be studied. The biggest concern is
  the aggressivity of $CF_4$ with some materials and $HF$ which can be formed by
  chemical reactions of $CF_4$ with water.

$\Diamond{}$ Optimize the detector configuration: 2-3 GEMs or 2 GEMs + MWPC.
  One of the most important requirements is a stable operation of the detector
  without sparking at a high enough gain to ensure the required electron identification.

$\Diamond{}$ Detector granularity needs to be studied and optimized. This is an important design 
  parameter affecting occupancy, hit pile-up, $DHR$ and last but not least cost. Our first 
  option here is to detect the blob in hexagonal pads of size comparable to the blob size. 
  This ensures as much charge as possible in one pad, a low gas gain factor in the 
  detector and low granularity. 

$\Diamond{}$ Study and optimize the HBD response to electrons versus hadrons. This is one of the
  most important features of the proposed detector. The electric field above the first GEM 
  plays a crucial role here~\cite{amos}. It affects the collection of photoelectrons 
  released from the $CsI$ and also the collected primary charge from the passage of a hadron.

$\Diamond{}$ Define the specifications and develop the front-end electronics. Analog information
  will be needed to recognize single and double hits. The specifications are closely linked 
  to other parameters of the system, like the pad size, the expected signal amplitude
  and shape. 

$\Diamond{}$ Study the influence of residual magnetic field on the pattern recognition.
  The second coil largely compensates the magnetic field created by the main coil. However 
  this does not result in a fully field-free region. Some residual field will be present
  and its influence on the pattern recognition has to be studied. 

$\Diamond{}$ Monte Carlo simulations. There is a broad program of simulations that have
  to be performed: (i) include the HBD in a full Monte Carlo simulation to study and
  optimize its response. (ii) How much material can be tolerated before the HBD?
  (iii) More realistic simulations of the HBD and its impact in rejecting the 
  combinatorial background, including all sources of electrons ($\gamma$ conversions, 
  Dalitz decays, resonance decays and semi-leptonic decays of open charm).

\vspace{1mm} 
Acknowledgments:
We acknowledge support from the Department of Energy and N.S.F. (USA), the US-Israel
Binational Science Foundation and the Israel Science Foundation.


\begin{thebibliography}{9}  
  
\bibitem{phnx} D.P. Morrison {\it et al.}, The PHENIX Collaboration,
  {Nucl. Phys. A 638, 565 (1998)}. For a comprehensive description of the PHENIX
  detector see the special NIM volume on the RHIC detectors, NIM (in press).
\bibitem{dsm} D. Mukhopadhyay (for the PHENIX Collaboration) To appear in the
  proceedings of Quark Matter 2002 (Nucl. Phys. A).
\bibitem{prop} Z. Fraenkel, B. Khachaturov, A. Kozlov, A. Milov,
  D. Mukhopadhyay, D. Pal, I. Ravinovich, I. Tserruya and S. Zhou, PHENIX
  Collaboration {Technical Note 391. http://www.phenix.bnl.gov/phenix/WWW/
    forms/info/view.html}.
\bibitem{hbd1} M. Chen {\it et al.}, {Nucl. Instr. and Meth. A346, 120 (1994)}. 
\bibitem{hbd2} R.P. Pisani, T.K. Hemmick, H.Chung, S.C. Johnson, T. Piazza,
  T. Vongpaseuth and M. Akopyan, {Nucl. Instr. and Meth. A400, 243
    (1997)}.
\bibitem{gem0} F. Sauli, {Nucl. Instr. and Meth. A386, 531 (1997)}.
\bibitem{amos} D. Mormann, A. Breskin, R. Chechik, P. Cwetanski, B.K. Singh
  {Nucl. Instr. and Meth. A478, 230 (2002)}, and R. Chechik, these proceedings.
\bibitem{cf41} HADES experiment, {Nucl. Instr. and Meth. A371, 300 (1996)}.
\bibitem{cf42} A. Pansky, A. Breskin, A. Buzulutskov, R. Chechik, V. Elkind and
  J. Va'vra, {Nucl. Instr. and Meth. A354, 262 (1995)}.
\bibitem{gem4} A. Breskin, A. Buzulutskov, R. Chechik, {Nucl. Instr. and
    Meth. A483, 670 (2002)}.
\bibitem{qe} F. Piuz, these proceedings and B.K. Singh, E. Shefer, A. Breskin,
  R. Chechik and N. Avraham, {Nucl. Instr. and Meth. A454, 364 (2000)}.
\end{thebibliography}
\end{document}